%% file: bare_conf.tex
\documentclass[conference]{IEEEtran}

\usepackage{cite}
\usepackage[pdftex]{graphicx}
\usepackage[cmex10]{amsmath}
\usepackage{algorithm}
\usepackage{algorithmic}
\usepackage{multirow}

\usepackage[caption=false,font=footnotesize]{subfig}

\input{Qcircuit.tex}

\begin{document}

\title{Synthesis of Topological Quantum Circuits}

\author{\IEEEauthorblockN{Alexandru Paler$^1$ \qquad Simon Devitt$^2$
    \qquad Kae Nemoto$^2$ \qquad Ilia Polian$^1$}\\[0.1cm]
  \IEEEauthorblockA{
  \begin{tabular}{c@{\hspace{2cm}}c}
    $^1$Faculty of Informatics and Mathematics & $^2$National Institute of Informatics \\
    University of Passau & 2-1-2 Hitotsubashi, Chiyoda-ku \\
    Innstr.~43, D-94032 Passau, Germany & Tokyo, Japan \\
    \{alexandru.paler$|$ilia.polian\}@uni-passau.de & \{devitt$|$nemoto\}@nii.ac.jp
  \end{tabular}
}}
\maketitle
\begin{abstract}
  Topological quantum computing has recently proven itself to be a very powerful model when considering large-scale, fully error corrected quantum architectures.  In addition to its robust nature under hardware errors, it is a software driven method of error corrected computation, with the hardware responsible for only creating a generic quantum resource (the topological lattice). Computation in this scheme is achieved by the geometric manipulation of holes (defects) within the lattice. Interactions between logical qubits (quantum gate operations) are implemented by using particular arrangements of the defects, such as braids and junctions. We demonstrate that junction-based topological quantum gates allow highly regular and structured implementation of large CNOT (controlled-not) gate networks, which ultimately form the basis of the error corrected primitives that must be used for an error corrected algorithm. We present a number of heuristics to optimise the area of the resulting structures and therefore the number of the required hardware resources.
\end{abstract}

\section{Introduction}

Quantum information science has been one of the extraordinary success stories of theoretical and experimental physics in the last 20 years. Since the introduction of the field in the 1980’s, there has been development of a complete theoretical framework for universal quantum computation and repeated experimental demonstration of quantum control on a small number of quantum bits (qubits) in multiple physical systems \cite{LJLNMO10}. In addition to the experimental system development, there have been multiple quantum architectures proposed demonstrating how a large-scale multi-million qubit machine could be constructed \cite{MTCCC05, FTYSPW07, DFSG08, MLFY10, JMFMKLY10, YJG10}. 

Even with this level of experimental development, large scale quantum information processing requires hardware that is extraordinarily accurate; many orders of magnitude higher than any system has ever demonstrated.  The inherent inaccuracies in quantum hardware induces errors during processing which much be corrected via dedicated and resource intensive Quantum Error Correcting (QEC) codes \cite{DMN09}. The topological model of QEC \cite{K03,DKLP02,RHG07} has shown itself to be more promising to many other long-standing techniques and currently forms the basis of effectively all modern quantum-computing architectures \cite{DFSG08, MLFY10, JMFMKLY10, YJG10}.  The continuing success of experimental fabrication now requires us to seriously address the programming and operation of a large scale computer. This work is aimed at developing the required synthesis tools for large scale quantum information processing in the context of a fully error corrected system.  We will present the classical framework for converting the abstract circuits describing a quantum algorithm into the full set of error corrected operations applied by the actual hardware. This work represents arguably the first serious attempt to develop a classical framework for a fully error corrected computational model compatible with current architectural designs.

Topological Quantum Computing (TQC) is a model for universal computation that is constructed on the framework of active error correction. This method of computation very abstract when compared to classical computer science.  One of the more bizarre aspects of this model is that   the physical hardware \textit{doesn't actually perform any real computation}. Instead, the hardware is only responsible for producing a very large three-dimensional lattice of qubits which are all linked together to form a single, massive quantum state. This quantum state forms the \textit{workbench} of the computation and information is created, processed and read-out via the strategic manipulation of this massive quantum state produced from the hardware \cite{FG08}.   The volume of this lattice directly relates to the physical resources required 
for an error corrected algorithm. As large quantum algorithms commonly require thousands of encoded qubits and high levels of error correction are often needed, the volume of the lattice is an important criteria for the synthesis of large quantum circuits. Minimising this volume reduces the amount of physical resources required, and lowers the costs of computation \cite{DSMN11+}. 

In the topological model, quantum gates are realised via geometric shapes, known as defects, moving through the lattice (e.g. see Fig.~\ref{fig:statedist}). These realisations enable highly regular arrangements which are easily optimised using discrete algorithms. Unlike in \cite{RHG07} where error corrected gates are realised by moving defects around each other (known as braiding), we utilise a technique where gates are realised through \emph{junctions} (where defects are jointed together in the lattice).  These constructions of error corrected gates are minimal with regard to the occupied lattice volume, and enable an efficient placement within it.

This work presents compact representations for the necessary TQC quantum gates, and uses these for several heuristics that minimise the size of the required qubit lattice, known as the topological cluster. To the best of our knowledge, this is the first automatic quantum synthesis procedure with fully error corrected, TQC as the target technology.

\section{Background}

\subsection{Quantum Computing}

Quantum computing can be, to a certain extent, described by building parallels to classical computing. A comprehensive 
review of quantum information and computation can be found in \cite{NC00}. The concept of classical bit has its quantum 
counterpart, called a quantum bit (\emph{qubit}). Like in classical computing, quantum gates are used to manipulate the state of a 
qubit, but in contrast to classical gates, quantum gates always have an equal number of input and output qubits. 

The state of a qubit $\psi$ can be represented as a vector $\ket{\psi}=\alpha_0\ket{0} + \alpha_1\ket{1}$, where $\alpha_0$ and $\alpha_1$ are complex numbers (called amplitudes) that satisfy $|\alpha_0|^2 + |\alpha_1|^2 = 1$. Similarly to the classical world, $\ket{0}$ and $\ket{1}$ are the states that are analogue to the classical bits $0$ and $1$. But unlike in classical computing, a qubit can be in a \emph{superposition} of these two states, where it is both 0 and 1. For example, the state $\alpha_0\ket{0}+\alpha_1\ket{1}$ has a probability of $|\alpha_0|^2$ of being measured in the $\ket{0}$ state and a probability of $|\alpha_1|^2$ of being measured in the $\ket{1}$ state. These  
states represent quantum wavefunctions and it is the interference of these wavefunctions  
driving quantum processing.  Algorithms are designed 
such that the amplitude related to the correct answer is amplified while the wrong answers are suppressed. 

Another similarity between quantum computing and classical computing is that computations can be expressed as circuits. \emph{Quantum circuits} are a description of a quantum algorithm (e.g. see Fig.~\ref{fig:teleport}). Quantum circuits have the input on the left and the output on the right. The horizontal \emph{wires} represent the qubits, and the computation is performed by applying a time ordered set of gates, left to right.

A quantum gate manipulating $m$ qubits is described by a $2^m \times 2^m$ unitary matrix acting 
on an column vector of length $2^{m}$ with entries $\{\alpha_i\}$ where $\sum_{i=0}^{2^m-1}|\alpha_i|^2 = 1$ 
for $i \in [0,..,2^{m}-1]$ representing 
the $m$-qubit state $\ket{\phi} = \sum_{i=0}^{m-1}\alpha_i\ket{i}$.  For example, the \emph{controlled-not} gate (CNOT) acting on two qubits is defined by the following $4 \times 4$ matrix:
\begin{equation}
  \label{eq:cnot}
  \textstyle CNOT = \left(
    \begin{array}{cccc}
      1 & 0 & 0 & 0 \\
      0 & 1 & 0 & 0 \\
      0 & 0 & 0 & 1 \\
      0 & 0 & 1 & 0
    \end{array}
\right).
\end{equation}

Performing the CNOT operation between two qubits $c$ and $t$ in some arbitrary quantum state\footnote{The state 
of the qubits is written in a shorthand notation: $\ket{c}\ket{t}=\ket{ct}$.} results in the following output:
\begin{IEEEeqnarray}{rCl}
	\ket{ct}_{in} & = & \alpha_{00}\ket{00} + \alpha_{01}\ket{01} + 
							\alpha_{10}\ket{10} + \alpha_{11}\ket{11} \nonumber \\
	\ket{ct}_{out} & = & \alpha_{00}\ket{00} + \alpha_{01}\ket{01} + 
							\alpha_{10}\ket{11} + \alpha_{11}\ket{10}
\end{IEEEeqnarray}

Although in this section only the single-target CNOT is discussed, the provided details can be easily generalised for the multi-target CNOT. In Fig.~\ref{fig:teleport} the CNOT gate is depicted as the vertical line connecting two qubits. The filled black dot represents the control qubit, while the $\oplus$ is the target qubit.

Reading the value of a bit has a quantum counterpart; measurement. Unlike classical readout, quantum measurement allows us to read out qubits in multiple ways.  The standard measurement in quantum computation, referred to as a $Z$-basis measurement, measures if the qubit is in the $\ket{0}$ or $\ket{1}$ state, collapsing (where the wavefunction describing the quantum state discontinuously 
changes) any superposition 
inconsistent with the measurement result. Another type of measurement is an $X$-basis 
measurement, which measures if the qubit is in the $\frac{1}{\sqrt{2}}\left(\ket{0}+\ket{1}\right)$ state or the $\frac{1}{\sqrt{2}}\left(\ket{0}-\ket{1}\right)$ state.  This type of measurement is valid as these two states are orthogonal (the wavefunctions describing these 
states have zero overlap).       
In Fig.~\ref{fig:teleport}, the encircled $X$ and $Z$ are representing these two types of measurements.

\subsection{Topological Quantum Computing}

Several quantum architectures are based upon the model of TQC \cite{DFSG08, MLFY10, JMFMKLY10}, as this method for computation has QEC integrated by construction. TQC is the preferred method for three primary reasons. 1) It has one of the highest fault-tolerant thresholds of any method of QEC\footnote{The Fault-tolerant threshold is the fundamental error rate associated with the physical hardware that must be reached before QEC becomes effective.}. 2) It is a \textit{local} model of computation, i.e. individual physical qubits in the computer only have to interact with their neighbours. 3) The quantum hardware is only used to prepare a large three-dimensional lattice of connected qubits (the topological cluster), algorithmic implementation is a function of how the lattice is consumed rather than how it is created. Therefore the TQC model is a software driven method of computation. 
 
\begin{figure}[t!]
	\centering
	\includegraphics[scale=0.3]{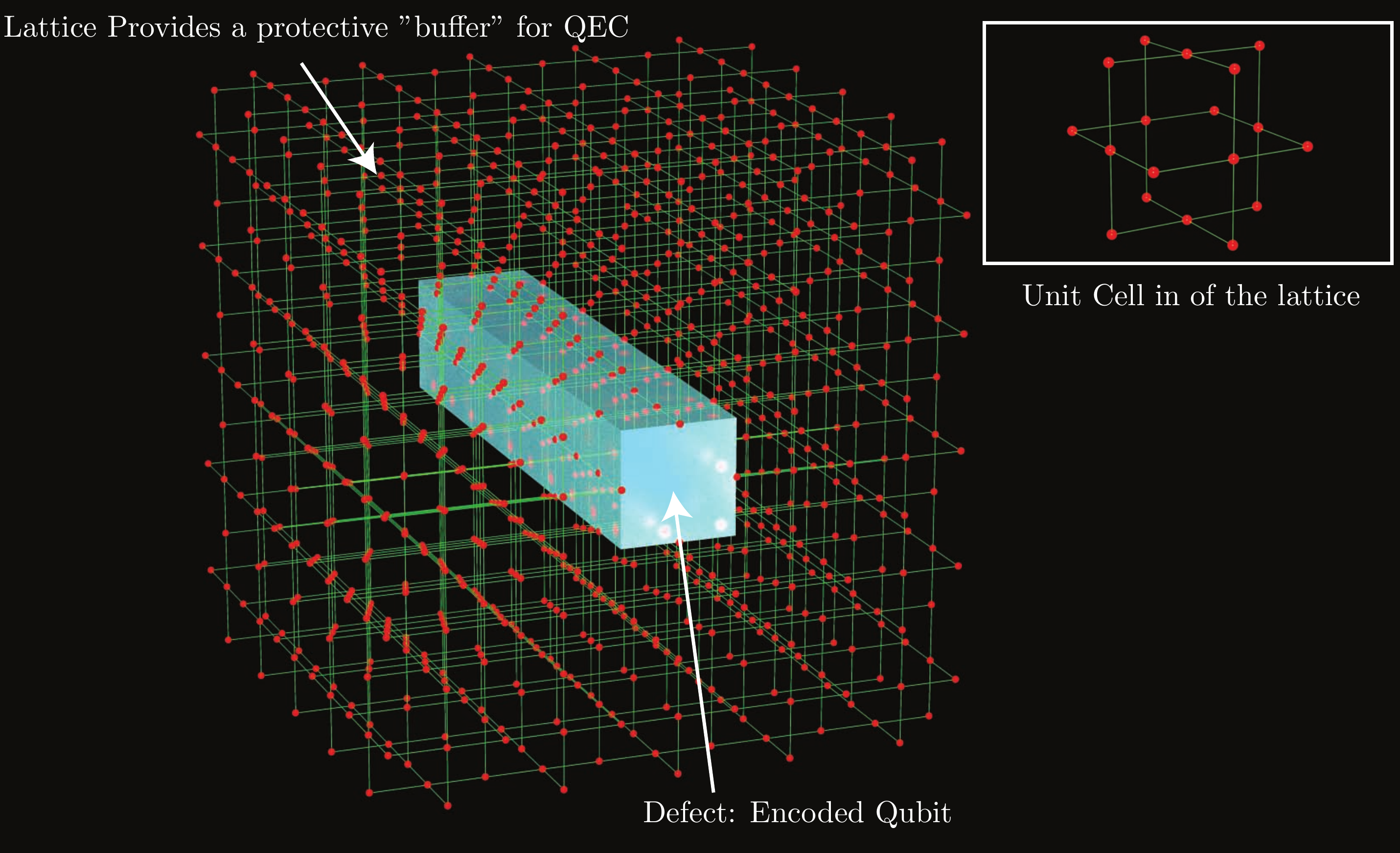}
	\caption{\textit{Logical volume of the topological cluster.} Each red dot represents a physical qubit, green lines represent links (entanglement between qubits). An encoded piece of information is a rectangular hole that is created by physically removing qubits internal to its boundary, the lattice surrounding the defect provides the error correction "buffer". The insert illustrates a unit cell of the lattice, consisting of 18 qubits.}
	\label{fig:logicalVol}
\end{figure}

The specifics of how TQC works is complicated, however the general principal can be explained.  The quantum hardware prepares a massive 3D lattice of qubits that are all connected (entangled) to form a single enormous quantum state.  The unit cell of this lattice is shown as the insert in Fig.~\ref{fig:logicalVol}.  Logical information is introduced and error protected by deliberately creating holes in this lattice.  Shown in Fig.~\ref{fig:logicalVol} is an example of a logically encoded qubit.  The information is stored within a hole (defect) in the lattice, this defect is created by simply removing the physical qubits internal to its boundary.  In Fig.~\ref{fig:logicalVol}, the defect is the rectangular structure and all physical qubits internal to this structure are simply removed from the lattice. The defect is surrounded by the bulk of the lattice, it is this bulk that provides the error protection of the model. If a defect has a large cross-section and is surrounded by a large "buffer" of the lattice, the information is heavily protected.  The details of how this works can be found in  \cite{RHG07,FG08}.

\begin{figure}[t!]
	\centering
	\includegraphics[scale=0.27]{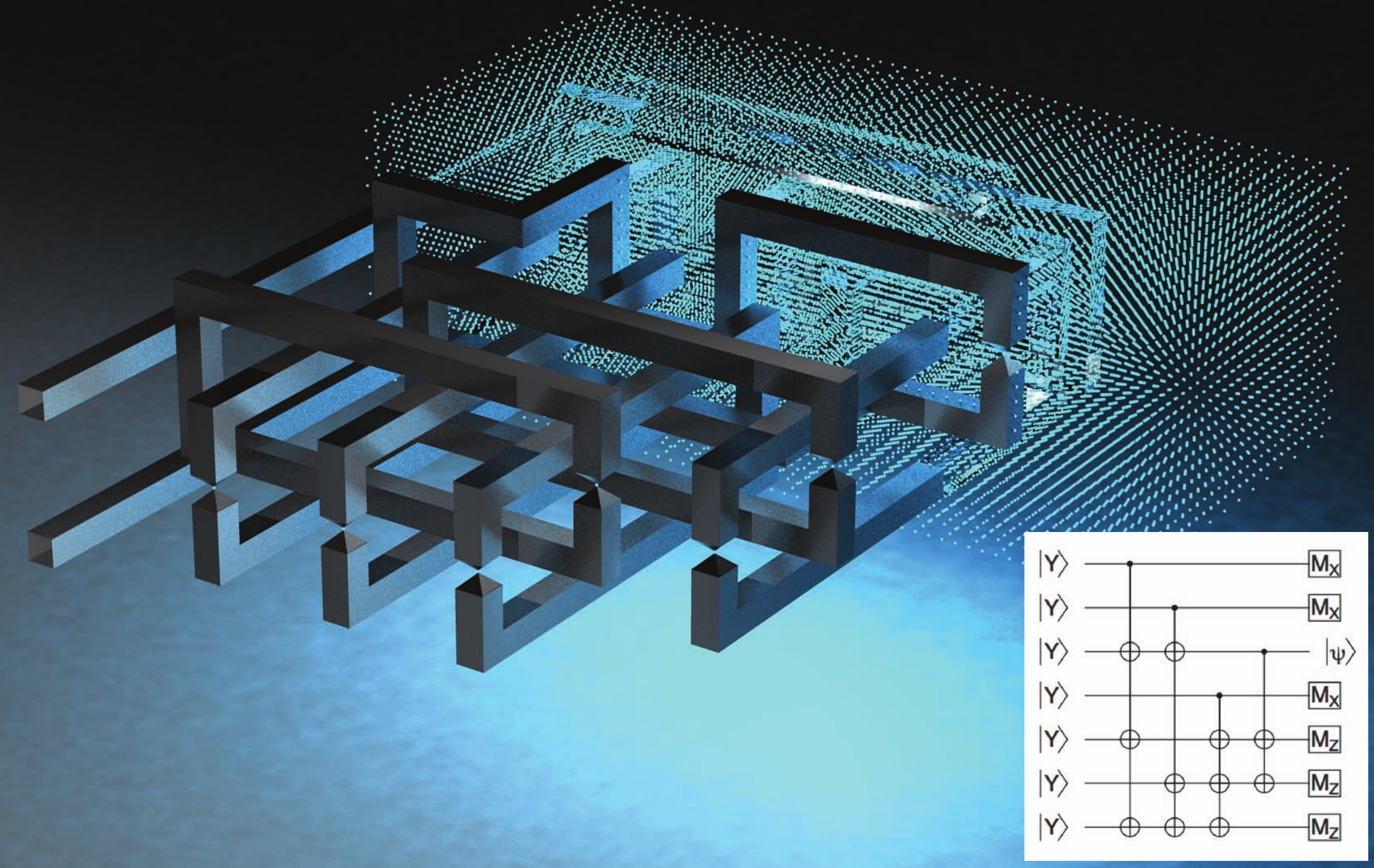}
	\caption{A topologically protected quantum circuit using defect braiding.  
	The insert illustrates the quantum circuit realised. CNOT gates are achieved via the geometric movement 
	of defect structures to form closed braids. The geometric 
	movement of defects is realised by the selective removal of physical qubits internal to the defect boundaries. The physical resources required are determined from the total volume of the lattice.}
	\label{fig:statedist}
\end{figure}

The purpose of computation with this model is to, in a controlled manner, manipulate the shape and movement of the defects within a large lattice produced by the hardware. Illustrated in Fig.~\ref{fig:statedist} is a larger quantum circuit (performing a sub-circuit called state distillation which is required to perform valid operations on encoded information). Instead of one defect, we have enough space to introduce multiple defects, representing multiple encoded qubits. These qubits are then interacted by moving defects around each other. These movements are performed by selecting which \textit{physical} qubits need to be removed from the lattice to form these structures.  

A large quantum algorithm is therefore defined via these geometric structures and how they are arranged and embedded within the topological cluster. As the total size of the lattice is directly related so the number of devices within our quantum computer, we clearly want to ensure that large quantum algorithms are classically compacted to occupy the smallest possible volume, reducing resources in the hardware. 

%\emph{topological quantum codes} such as the surface code \cite{NJP}
%and the Raussendorff code \cite{PRA}. The error-rate thresholds of
%these codes are at levels which can be expected from most advanced
%experimental systems in the near future. In TQC, it is distinguished
%between \emph{logical qubits} and \emph{physical qubits}. Logical
%qubits represent quantum information.

%The physical qubits in TQC are physical devices which have all the
%properties of a regular qubit and can assume superposition states. In
%today's experiments, quantum optical systems are used to obtain
%physical qubits but other technologies could be employed as well.
%Physical qubits are arranged to form a \emph{topological cluster},
%which consists of two interlaced three-dimensional lattices, known as
%the primal and dual spaces, respectively. Logical qubits are
%implemented by holes in these lattices. An individual logic qubit is
%represented by a pair of holes in either the primal or the dual space.
%Operations between logic qubits are performed by positioning the
%respective holes in specific patterns. Braiding a hole from the primal
%space with one from the dual space induces a CNOT operation between
%the logical qubits represented by these holes \cite{??}. CNOT
%operations between two holes in the primal space can be implemented by
%several braids involving auxilliary holes from the dual space
%representing what is known as ancilla qubits. 

\section{Gates for Topological Quantum Computing}

Universal quantum computation must be performed by using a discrete and reduced set of gates (whose 
direct application is compatible with the underlying encoding). In the context of TQC the chosen universal gate set consists of the CNOT gate and a discrete set of single qubit gates \cite{BK05+,DN06}.

Single-qubit gates are applied via teleportations to a data qubit, by utilising an ancilla (auxiliary) qubit that is prepared in a specific encoded state. This preparation has to be further decomposed into distillation protocols which act to purify multiple noisy copies of these auxiliary states (as they are introduced/injected into the lattice in a non fault-tolerant manner) into a single highly pure copy that can be coupled without introducing additional noise into our encoded data. 
A more in-depth discussion of how universality is achieved for TQC is given in \cite{FG08}.

State injection and distillation will not be addressed in this work, because they don't dictate the general framework of the synthesis. Fully error corrected computations are ultimately a large array of CNOTs and identities within the cluster. The synthesis of computation will not affect the universality of the TQC model.

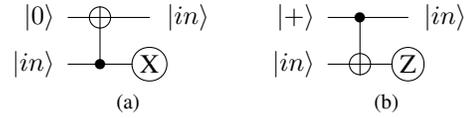
\begin{figure}[t!]
\centerline{
\subfloat[]{\label{fig:teleporta}\Qcircuit @C=0.8em @R=.5em @!R{
	   \lstick{\ket{0}} & \targ & \qw & \ket{in}\\
	 	 \lstick{\ket{in}} & \ctrl{-1} & \measure{\mbox{X}}
  }}
\hfil
\subfloat[]{\label{fig:teleportb}\Qcircuit @C=0.8em @R=.5em @!R{
	   \lstick{\ket{+}} & \ctrl{1} & \qw & \ket{in}\\
	   \lstick{\ket{in}} & \targ & \measure{\mbox{Z}}
  }}
}
\caption{Quantum teleportation circuits: The arbitrary $\ket{in}$ state is teleported from the lower qubit to the upper qubit.}
\label{fig:teleport}
\end{figure}

\subsection{The Topological CNOT Gate}

%The CNOT gate is one of the most used gates, but is a conceptually difficult operation to understand in the TQC model. In this section we will simply use the TQC implementations of the teleportation circuits shown in Fig~\ref{fig:teleport} \cite{RHG07} as primitives for constructing the gate (see Fig.~\ref{fig:raussendorfjunctions}).

%It is known how to perform teleportations in the TQC model, and it is also known that a teleportation consists of a CNOT and a measurement. 
%However, the topological teleportation is an atomic operation, and it is impossible to separate it into its constituents. 
%Regardless of the inputs sent to a teleportation circuit, its structure remains the same, even if its outputs will not represent a teleported 
%state. For these reasons, the topological teleportation implementations are used for constructing a topological CNOT, but the inputs will be 
%selected so that the outputs will be exactly those of an \emph{emulated} CNOT.

\begin{figure}[t!]
	\centerline{
		\subfloat[]{\begin{minipage}{0.4\linewidth}
			\includegraphics[scale=1.5]{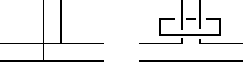}
		\end{minipage}
		}
		\hfill
		\subfloat[]{\label{fig:raussa}
		\begin{minipage}{0.2\linewidth}
			\Qcircuit @C=0.8em @R=1.1em @!R{
	   		 & \ctrl{1} & \qw \\
	   		\lstick{\ket{0}} & \targ & \qw
  			}
		\end{minipage}
		}
		\hfill
		\subfloat[]{\label{fig:raussb}
		\begin{minipage}{0.2\linewidth}
			\Qcircuit @C=0.8em @R=.5em @!R{
	   		 & \ctrl{1} & \measure{\mbox{X}}\\
	   		 & \targ & \qw
  			}		
		\end{minipage}
		}
	}
	\caption{Teleportation circuits as junctions. Encoded qubits are represented as pairs of defects, CNOT operations, initialisation and measurement occur at the junction point \cite{RHG07}.  A defect ring is 
	required in Fig. a) to ensure correct operation.  This ring is illustrated more clearly in Fig. \ref{fig:cnotcross}.}
	\label{fig:raussendorfjunctions}
\end{figure}

The CNOT gate is one of the most used gates, however as complete derivation of the junction CNOT in this work is rather complicated, we instead utilise the teleportation circuits introduced in \cite{RHG07} as an assumed starting point. These defect junctions invoke the circuits illustrated in Fig.~\ref{fig:raussendorfjunctions}, where each encoded qubit is represented by a pair of defects (holes in the lattice). In one case (Fig.~\ref{fig:raussa}), a CNOT operation is performed between the input qubit and a freshly initialised qubit in the $\ket{0}$ state, with both outputted.  In Fig.~\ref{fig:raussb} two encoded qubits are used as input and at the junction point, a CNOT operation is performed, then one of the qubits is measured in the $X$ basis.  Before the junction, a ring encircles the two defects and 
is required to ensure that the logical information is 
propagated correctly according to the quantum circuit. 
These circuits are generalisations of the teleportation circuits in Fig.~\ref{fig:teleport} \cite{RHG07} 
and have a similar geometric structure.

Given these two junctions and their correspondence to the associated quantum circuits, the derivation of the CNOT is achieved by simply attaching one of the output qubits from circuit one to one of the inputs of circuit two and calculating the resulting circuit identity. 
\begin{IEEEeqnarray}{rCl}
	\ket{c0t} & = & 
a_0\ket{000} + a_1\ket{001} + a_2\ket{100} + a_3\ket{101} \nonumber\\
\text{CNOT}	& \rightarrow & 
a_0\ket{000} + a_1\ket{001} + a_2\ket{111} + a_3\ket{110} \nonumber\\
\text{Measure 2 in } X	& \rightarrow & 
a_0\ket{00} + a_1\ket{01} \pm a_2\ket{11} \pm a_3\ket{10} 
 \label{eq:cnot2}
\end{IEEEeqnarray}

where the $\pm$ signs depend on the measurement result of qubit two and can be corrected.  

\subsection{The Identity Gate}

Another gate that can be directly implemented is the identity gate.  This gate is simply reshaping the geometric structure of the defects throughout the lattice. If a pair of defects are utilised to support a single encoded qubit of information then movements must be performed in a manner as to maintain a constant separation of the two defects (this is to ensure that the error correction properties of the code are maintained).  Geometrically, defects may be moved in any directions in the 3D lattice provided there is sufficient bulk space\footnote{Bulk space in this context means part of the lattice \textit{not} containing other defects.}.

\section{Synthesis for TQC}

TQC synthesis is the process by which a quantum computation (e.g. described using the high-level quantum circuit formalism) is translated into a representation of topological quantum gates. The synthesis process is conceptually similar to classical logical synthesis. The CNOT and the identity can be regarded as the building-blocks necessary for TQC, because as arbitrary single-qubit gates are constructed using CNOT gates, qubit initialisation and measurement in two orthogonal bases (initialisation of encoded qubits are realised by introducing defects through 
the removal of physical qubits while measurement is realised by terminating a defect by no longer removing 
physical qubits from the lattice).

Note that the problem formulation is different from the synthesis problem in the field of reversible circuits (Boolean circuits 
consisting of CNOT gates and other bijective gates). Reversible circuit synthesis \cite{shende2003synthesis, patel2004fault, 
saeedi2011synthesis, gupta2006algorithm, wille2009bdd} takes either a high-level description or a gate-level net-list of a circuit as 
the input, and calculates an (optimized) gate-level net-list with an equivalent functionality. In contrast, the presented algorithms 
generate the arrangement of \emph{defects} of a topological cluster. No optimisations of the gate-level net-list is performed, 
although such optimisations could further improve the outcome of the algorithms. The major difference between reversible computation 
and TQC is that, although both make extensive use of the CNOT gate, only in TQC do we require state injection and distillation for 
achieving the \emph{complete} set of universal gates. Therefore, TQC can implement arbitrary quantum circuits which process quantum 
states incorporating superposition and entanglement. Reversible circuits process Boolean states, i.e., vectors of 0s and 1s, and are 
complete within this paradigm.

%Reversible circuits are only able to represent the reduced set of Boolean states.

A quantum computation can be expressed as its constituent set of gates $G$, and, therefore, for TQC synthesis, $G$ will consist only of CNOT gates. Assuming that all the qubits of a computation are numbered ($Q$ is the set of qubits), a multi-target CNOT $cn \in G$ will be characterised by a qubit $control(cn) \in Q$ indicating the control, and a set of qubits $targets(cn)$ indicating the targets.

\begin{figure}[!t]
\centering
\begin{minipage}{\linewidth}
	\includegraphics[scale=1.2]{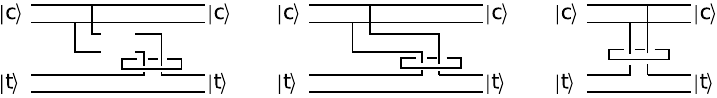}
	\hfill
\end{minipage}
\begin{minipage}{\linewidth}
	\hfill
	\vspace{0.2cm}
\end{minipage}
\\
\hfill
\begin{minipage}{0.4\linewidth}
%  \Qcircuit @C=1.3em @R=.7em @!R{
  \Qcircuit @C=1.3em @R=.7em{
   \lstick{\ket{c}} & \qw &\ctrl{1} & \qw & \qw & \qw & \ket{c}\\
   & \lstick{\ket{0}} & \targ & \qw & \ctrl{1} & \measure{\mbox{X}} & &\\
   \lstick{\ket{t}} & \qw & \qw & \qw & \targ & \qw & \ket{c \oplus t} 
\gategroup{1}{1}{2}{3}{0.7em}{--}\gategroup{2}{4}{3}{6}{0.7em}{--}
  }
\end{minipage}
$\rightarrow$
\hfill
\begin{minipage}{0.4\linewidth}
%  \Qcircuit @C=1em @R=.7em @!R{
  \Qcircuit @C=1em @R=.7em{
   \lstick{\ket{c}} & \ctrl{1} & \qw & \qw & \ket{c}\\
   \lstick{\ket{0}} & \targ & \ctrl{1} & \qw & \measure{\mbox{X}} &\\
   \lstick{\ket{t}} & \qw & \targ\ & \qw & \ket{c \oplus t}
  }
\end{minipage}
\hfill
\caption{The topological CNOT and its quantum circuit equivalent after combining the two teleportation circuits from Fig.~\ref{fig:raussendorfjunctions}}
\label{fig:combined}
\end{figure}

The structure of the topological QEC is such that encoded qubits form a three-dimensional algorithmic layout. This layout allows nearest neighbour (where encoded qubits only interact with their neighbours) or long range interactions between separate regions of the computer. Utilising long range interactions will have an impact on the ability to optimise the total volume of cluster required for a large computation. Nearest neighbour interactions can be easily implemented by translating quantum algorithms into a two-dimensional field, having as a result a reduced number of possibilities in which a defect can be moved through the cluster. For an operation between two neighbouring encoded qubits to take place, the qubits will be firstly brought near to each other in the field, and will interact afterwards. In this work, in the context of TQC synthesis, the area of the two-dimensional field is the cost function to minimise.

\subsection{Compact Representation of Gates}

%The gates necessary for TQC can be transformed to the two-dimensional representations used in this formalism, as illustrated in Fig.~\ref{fig:cnotcross}. The first 6 (4 \emph{corners}, 1 horizontal and 1 vertical) representations are for the identity gate acting on a single qubit (a single qubit is encoded using two defects). The seventh identity gate illustrated appears in the two-dimensional representation to interact, but in the actual 3D structure, one pair of defects passes under the other.  The CNOT gate is almost identical to the identity, but has a discontinuous qubit as the target which is physically connected to the control. From a two-dimensional view, the control 

\begin{figure}[t!]
	\centering
	\includegraphics[scale=0.28]{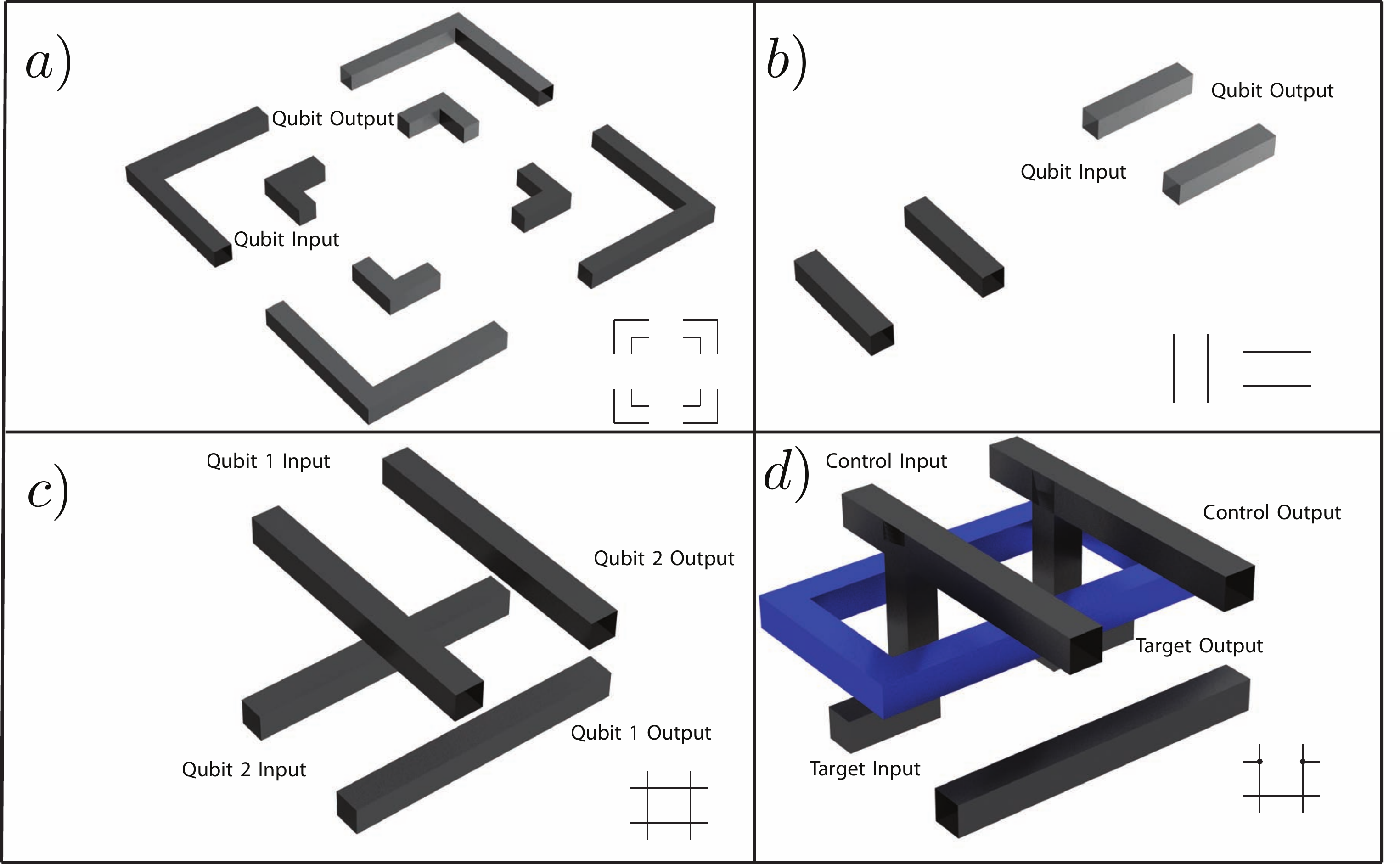}
	\caption{Compact topological representations of the identity gates and of the CNOT.  Each encoded qubit is a pair of defects which always maintain a fixed separation from all other defects (except at junction points), this maintains the error correction strength of the code.  The defect ring that encircles the CNOT is implicit in the 
	2D representation.}
	\label{fig:cnotcross}
\end{figure}

The TQC synthesis algorithm uses eight primitives (see Fig.~\ref{fig:cnotcross}) to implement the two gates: identity and CNOT. Both the 3D realisation and the two-dimensional representation of the primitives are shown in the figure.  Recall that each qubit is represented by a pair of defects running in parallel. The six primitives from Fig.~\ref{fig:cnotcross}a,b perform the identity operation on one qubit (keep one qubit unchanged). The primitive in Fig.~\ref{fig:cnotcross}c keeps two qubits (represented by horizontal and vertical pairs of defects) unchanged. Note that these two pairs of defects appear to touch each other in the two-dimensional representation, but in 3D they pass under each other. From a 3D perspective, the CNOT (Fig.~\ref{fig:cnotcross}d) has a discontinuous qubit as the target which is physically connected to the control. In two dimensions, the control qubit is vertical and the target qubit is horizontal.  The ring 
that encircles the junction within the CNOT is implicit within the 2D representation.

There is a direct mapping from the two-dimensional representation to the topological cluster, and assuming that for a computation all CNOTs and identities are placed on the same \emph{layer}, then the compact two-dimensional representation is adequate for representing arbitrary circuits.

\subsection{The Field of CNOTs}

The field of CNOTs is a matrix-like regular structure having the following properties:
\begin{itemize}
	\item the target qubits of each gate are located on the rows;
	\item the control qubits of each gate are located on the columns.
\end{itemize}
This arrangement is possible, because in 3D, the planes supporting the qubits are parallel (see Fig.~\ref{fig:cnotcross}).

The area of the field depends on the width (number of columns) and the height (number of rows), that in turn are influenced by 
characteristics of the quantum computation, such as the number of gate and qubits. Throughout the following sections a distinction 
will be made between cluster (a 3D structure with fixed depth) and two-dimensional field. Also, because the cluster volume is 
linearly related to the field area, when analysing the efficiency of the synthesis, the area will be referred to.

\begin{figure}[t!]
\centerline{
\subfloat[]{\label{fig:fexamplea}\begin{minipage}{0.5\linewidth}\Qcircuit @C=1.0em @R=.5em @!R{
			& cn_1 & cn_2 & cn_3 &&\\
			\lstick{\ket{q_0^i}} & \targ \qwx[3]& \control \qw & \qw & \qw & \ket{q_0^o}\\
			\lstick{\ket{q_1^i}} & \targ & \qw & \targ \qwx[1]& \qw & \ket{q_1^o}\\
			\lstick{\ket{q_2^i}} & \qw & \qw & \control \qw & \qw & \ket{q_2^o}\\
			\lstick{\ket{q_3^i}} & \control \qw & \targ \qwx[-3]& \qw & \qw & \ket{q_3^o}
			\gategroup{2}{2}{5}{2}{0.5em}{--}
			\gategroup{2}{3}{5}{3}{0.5em}{--}
			\gategroup{3}{4}{4}{4}{0.5em}{--}
		}\end{minipage}}
\hfil%
\subfloat[]{\label{fig:fexampleb}\begin{minipage}{0.2\linewidth}\includegraphics[scale=1]{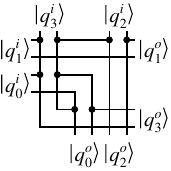}\end{minipage}}%
\hfil%
\subfloat[]{\label{fig:fexamplec}\begin{minipage}{0.3\linewidth}\includegraphics[scale=1]{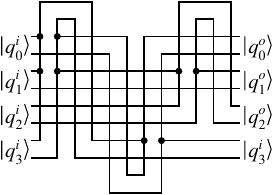}\end{minipage}}%
}
\caption{A circuit and two possibilities to construct the corresponding fields}
\label{fig:fexample}
\end{figure}

Two alternative two-dimensional fields implementing the 4-qubit, 3-CNOT circuit from Fig.~\ref{fig:fexamplea} are shown in 
Fig.~\ref{fig:fexampleb} and \ref{fig:fexamplec}. Note that $cn_1$ is a multi-target CNOT (it has two targets) and is represented by 
two single-target CNOTs. The $cn_2$ operation is implemented on the second column of the field: qubit $q_0$ is transformed into a 
control for qubit $q_3$. Finally, $cn_3$ is implemented on the third column. Positions not occupied by CNOTs implement identities. 
The area is $3 \times 3 = 9$. The field in Fig.~\ref{fig:fexamplec} implements the same computation in a less efficient way using 
$12$ identity gates, and has an area of $8 \times 6 = 48$.

%The first field has 3 rows and 3 columns. The $(0,0)$ and $(1,0)$ cells are representing the individual single-gate 
%CNOTs that are the constituents of the first multi-target CNOT of the example circuit. As a result, the first 
%column of the field contains the first CNOT. 

%The second column of the field contains $cn_2$, but the $cn_(1,1)$ cell contains a 
%single-qubit identity gate, that is transforming the previous target qubit $q_0$ into a control qubit. The $(2,2)$ cell is representing the 
%identity between the $q_2$ and $q_3$ qubits. The inputs of the field are placed on the left and at the top, the outputs are placed on the 
%right and at the bottom. The first field has an area of $3 \times 3 = 9$.

%The second field has 8 rows and 6 columns, but it depicts the same quantum computation of 3 CNOTs. In this case the inputs to the field are 
%on the left, and the outputs are on the right of the field, and there are $12$ two-qubit identity gates used (compared to $1$ in the previous 
%field). The second field has an area of $8 \times 6 = 48$.

\subsection{Synthesis Algorithms}

The synthesis of quantum computations for a TQC cluster can be automated, and, as previously mentioned, an important aspect is the 
area of the synthesised field. In this section two synthesis algorithms are presented and evaluated, and both algorithms receive the 
set of gates $G$ as input, and output a field of CNOTs. The algorithms were designed starting from the assumption, that the output 
field (topological cluster) will be \emph{consumed} from the left to the right by the TQC hardware. Therefore, the first gate will be 
the left-most in the field. The computational complexity of the algorithms is linear in the number of gates.

Algorithm~\ref{alg:alg1} is constructing the field on a per-gate basis, and inserts qubits involved in the computation as they are 
needed. If a target qubit does not exist in the circuit created so far, it is inserted on a row, and if a control qubit is needed, 
this will be inserted directly on a column. Also, if for a given gate, that is currently synthesised, an existing target qubit will 
be used as a control, then the qubit will be moved to the next free column. An existing control qubit needed as a target, will be 
first moved to the next free row, and then to the next free column. 

\begin{algorithm}
	\caption{Unbounded: Moving the control downwards}
	\label{alg:alg1}
	\begin{algorithmic}[1]
		\STATE Initialise an empty field
		\STATE{Start with no qubits inserted in the field}
		\FORALL{CNOT $cn \in G$}
			\FORALL{$q \in targets(cn) \cup \{control(cn)\}$}
				\IF{$q$ does not exist in the field}
					\STATE insert a row/column for $q$
				\ELSE
					\STATE move qubit $q$ to the next free row
				\ENDIF	
			\ENDFOR
			\STATE move all $q \in targets(cn) \cup \{control(cn)\}$ to the next free column
			\STATE move $control(cn)$ to the next free row using it as control for $targets(cn)$
		\ENDFOR
	\end{algorithmic}
\end{algorithm}

The fields generated by Algorithm~\ref{alg:alg1} typically have the main diagonal occupied, leading to a suboptimal area utilization. With regard to the field area, the worst-case scenario is achieved when the control of each gate was a control, and the number of targets equals the maximum number of qubits: $A_1^w =  |G|^2 \cdot (|Q| - 1) \cdot |Q|$.

A second algorithm was designed after noticing the worst-case negative impact of the unbounded construction of the field. If $G$, the set of gates, is analysed before the synthesis, useful information can be extracted, such as the number of qubits $|Q|$ involved in the quantum computation. The information can be used to initialise the field, thus bounding the number of rows, and also leading to a more efficient and predictable placement of the gates.

\begin{algorithm}
	\caption{Bounded: Weaving of qubits}
	\label{alg:alg2}
	\begin{algorithmic}[1]
		\STATE Compute $|Q|$ from $G$
		\STATE Initialise a field with $|Q| + 2$ rows
		\STATE Place each qubit $q_i \in Q $ on the row $i+1$
		\FORALL{CNOT $cn \in G$}
			\STATE Select the target qubit $control(cn)$
			\STATE Transform $control(cn)$ into a control qubit
			\STATE Continue $control(cn)$ while using it as a control for $targets(cn)$: a) downwards to row $|Q| + 1$; b) to the 
right one column;	c) upwards to row $0$; d) to the right one column; e) downwards to row $control(cn) + 1$.
			\STATE Transform $control(cn)$ back into a target qubit
		\ENDFOR
	\end{algorithmic}
\end{algorithm}

The output of Algorithm~\ref{alg:alg2} resembles a \emph{fabric of qubits}, because when $control(cn)$ is transformed into a control and back into a target, the control qubit is moved from a row to a column and back. The transformations are equivalent to the movements of a qubit inside a TQC cluster (see Fig.~\ref{fig:cnotcross}), and all the possible targets are affected because of the control qubit movement pattern.

The algorithm requires two \emph{buffer} rows (the highest and the lowest) to allow the control qubit to change its direction from upwards to downwards, and/or the other way around. Because a complete movement of the control qubit \emph{occupies} three columns of the field, the worst-case area of the output field will be $A_2^w = 3 \cdot (|Q| + 2) \cdot |G|$. However, the area of the output field can be improved by incorporating additional heuristics into the algorithm. For example, information extracted from the set of targeted qubits could be used for the control qubit movement. If, for a given CNOT $cn$, all the target qubits are placed above the control qubit ($\forall t \in G, control(cn) > t$), then the control will be only moved upwards in the field. The downwards movement is triggered when all the targets are placed below the control ($\forall t\in G, control(cn) < t$).

\section{Simulation Results}

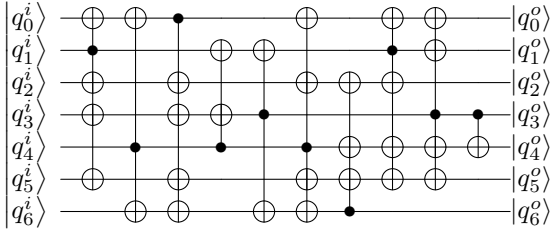
\begin{figure}[t!]
	\hfil
	\noindent\begin{minipage}{0.9\linewidth}
		\input{imgs/example.tex}
	\end{minipage}
	\caption{An sample circuit used for the bounded and the unbounded synthesis}
	\label{fig:algexampleqc}
\end{figure}

\begin{figure}[t!]
	\includegraphics[scale=1.0]{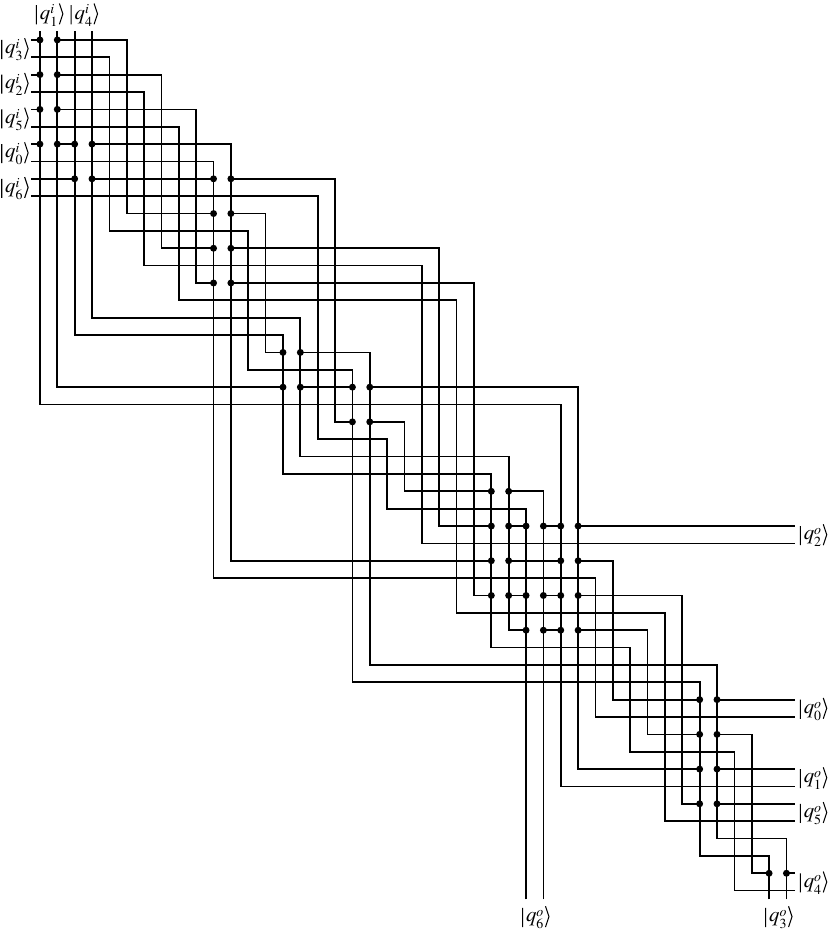}
	\caption{Output field of the unbounded synthesis}
	\label{fig:algexample1}
\end{figure}

\begin{figure}[t!]
	\includegraphics[scale=1.0]{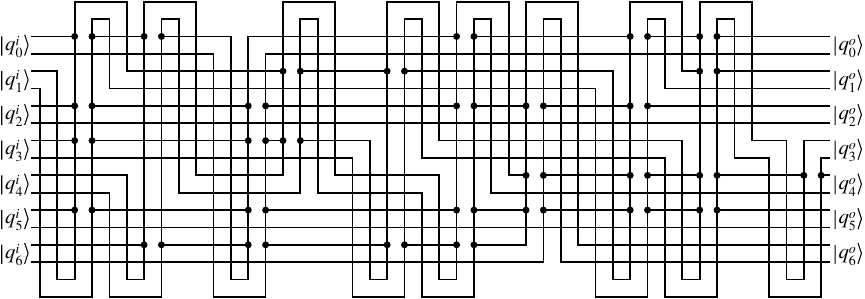}
	\caption{Output field of the bounded synthesis}
	\label{fig:algexample2}
\end{figure}

The presented algorithms were implemented, and using Fig.~\ref{fig:algexample1} and Fig.~\ref{fig:algexample2} the typical output fields generated by each algorithm can be visually compared. The quantum computation for which the fields were synthesised is depicted in Fig.~\ref{fig:algexampleqc}: it is applied on 7 qubits, and consists of 10 CNOTs targeting maximum 4 qubits.

For the evaluation of the algorithms, randomly generated circuits consisting entirely of CNOTs were used. The simulation results are presented in Table~\ref{tab:alg1}. The simulations were parametrised for $10$, $100$, $1000$ and $10000$ gates operating on $10$, $100$ and $1000$ qubits. The \emph{MT} column contains the value of the maximum size of $targets(cn)$ per randomly generated CNOT, and the \emph{Columns} and \emph{Rows} columns contain the average number of rows and columns obtained by the algorithms after executing them $100$ times for each combination of simulation parameters. Algorithm~\ref{alg:alg2} always generates a field with $|G| + 2$ rows, therefore column \emph{Rows} has been omitted from the table. It can be observed that the $MT$ parameter is directly influencing the average number of columns and rows obtained through Algorithm~\ref{alg:alg1}, but the results of Algorithm~\ref{alg:alg2} are not as strongly affected. 

The \emph{Red} column contains the reduction obtained when utilising the bounded synthesis instead of the unbounded algorithms. The unbounded algorithm can also perform better than the bounded algorithm (an example is shown in Figs.~\ref{fig:fexampleb} and \ref{fig:fexamplec}). The simulation results contain such a situation: circuits with $100$ qubits and $10$ CNOTs having a maximum of $20$ targets (see the highlighted entry in Table~\ref{tab:alg1}). In general, unbounded synthesis could perform better for circuits with less gates 
and less targets per CNOT.

Overall, the average area of the fields synthesised by the unbounded algorithm is larger compared to the average field area obtained with the bounded synthesis. This effect is particularly significant for large circuits like the ones consisting of $10000$ CNOTs.

\input{tables.tex}

%The layout of the output fields can be also applied as a criteria for comparing the algorithms. There are often situations, when a 
%synthesised computation will be repeatedly used as a subroutine of a larger computation (similar to a classical semi-custom design). The 
%bounded algorithm will deliver \emph{better} results, because all the input qubits and the output qubits are placed on the same row, 
%enabling a more convenient way of connection subroutines.

%There are situations, too, when the unbounded algorithm performs better then the bounded: for small circuits with a reduced maximum number 
%of targets per CNOT. An example is offered in Fig.~\ref{fig:fexample}. Nevertheless, an advantage of both algorithms is that 
%the area of the output field can be computed beforehand, because of their deterministic functionality. This fact can be useful when having to 
%decide which algorithm to use.

The execution times of the algorithms were negligible, and since the worst-case area consumption can be computed before running the algorithm, it is easier to decide for a given circuit which algorithm to use. An advantage of the bounded algorithm is that input and output qubits are placed on the same row. This is useful when synthesising sub-circuits of a larger circuit, as the subsequent connections of the sub-circuits may be simpler.

\section{Conclusion}

The integration of large-scale quantum algorithms with the strict requirements of fault-tolerant, error corrected quantum computation is a neglected area of research. While much research has been performed looking at methods for optimising quantum algorithms, these efforts did not take into consideration the requirements and restrictions of error corrected computing such as TQC.
The formulation of synthesis for TQC was presented, the primitives were constructed and two algorithms were designed. An important resource of TQC is the volume of the lattice needed for computation, and hence the algorithms were evaluated with respect to circuit volumes.

Future work will consist on improving the efficiency of the synthesis. For example, irregular (not rectangular) empty cluster space should be efficiently occupied.  The properties of the cluster and how topological circuits are constructed offer significant options for optimisation which are still currently under development. 

\section*{Acknowledgements}
SJD and KN are partially supported by the Quantum Cybernetics (MEXT) and FIRST projects, Japan.



\bibliography{paper}
\bibliographystyle{IEEEtran}

\end{document}

%% file: imgs/example.tex
\Qcircuit @C=0.8em @R=.4em @!R{
\lstick{\ket{q_0^i}} & \targ \qwx[1]& \targ \qwx[4]& \control \qw & \qw & \qw & \targ \qwx[4]& \qw & \targ \qwx[1]& \targ \qwx[3]& \qw & \qw & \ket{q_0^o} \\
\lstick{\ket{q_1^i}} & \control \qw & \qw & \qw & \targ \qwx[3]& \targ \qwx[2]& \qw & \qw & \control \qw & \targ & \qw & \qw & \ket{q_1^o}\\
\lstick{\ket{q_2^i}} & \targ & \qw & \targ & \qw & \qw & \targ & \targ \qwx[4]& \targ & \qw & \qw & \qw & \ket{q_2^o}\\
\lstick{\ket{q_3^i}} & \targ & \qw & \targ & \targ & \control \qw & \qw & \qw & \qw & \control \qw & \control \qw & \qw & \ket{q_3^o}\\
\lstick{\ket{q_4^i}} & \qw & \control \qw & \qw & \control \qw & \qw & \control \qw & \targ & \targ & \targ & \targ \qwx[-1]& \qw & \ket{q_4^o}\\
\lstick{\ket{q_5^i}} & \targ \qwx[-4]& \qw & \targ & \qw & \qw & \targ & \targ & \targ \qwx[-4]& \targ \qwx[-2]& \qw & \qw & \ket{q_5^o}\\
\lstick{\ket{q_6^i}} & \qw & \targ \qwx[-2]& \targ \qwx[-6]& \qw & \targ \qwx[-3]& \targ \qwx[-2]& \control \qw & \qw & \qw & \qw & \qw & \ket{q_6^o}
}

%% file: tables.tex
\begin{table}
\renewcommand{\arraystretch}{1.1}
\caption{Unbounded and Bounded Synthesis Results}
\label{tab:alg1}
\centering
\begin{tabular}{@{\hspace{0.05cm}}c@{\hspace{0.05cm}}@{\hspace{0.05cm}}c@{\hspace{0.05cm}}@{\hspace{0.05cm}}c@{\hspace{0.05cm}}|@{\hspace{0.05cm}
}c@{\hspace{0.05cm}}@{\hspace{0.05cm}}c@{\hspace{0.05cm}}@{\hspace{0.05cm}}c@{\hspace{0.05cm}}|@{\hspace{0.05cm}}c@{\hspace{0.05cm}}@{\hspace{0.1
cm}}c@{\hspace{0.05cm}}|@{\hspace{0.05cm}}c@{\hspace{0.05cm}}}
    \hline
		\multirow{2}{*}{$|$Q$|$} & \multirow{2}{*}{$|$G$|$} & \multirow{2}{*}{MT} & 
\multicolumn{3}{@{\hspace{0.05cm}}c@{\hspace{0.05cm}}|@{\hspace{0.05cm}}}{Algorithm 1} & 
\multicolumn{2}{@{\hspace{0.05cm}}c@{\hspace{0.05cm}}|@{\hspace{0.05cm}}}{Algorithm 2} & 
\multirow{2}{*}{Red}\\
     & & & Rows & Cols & Area & Cols & Area & \\
    \hline
    \hline
10 & 10 & 2 & 18 & 13 & 2.7e+02 & 17 & 1.9e+02 & 1.42e+00\\
10 & 10 & 5 & 33 & 19 & 6.2e+02 & 18 & 2.1e+02 & 2.95e+00\\
10 & 10 & 9 & 57 & 30 & 1.7e+03 & 19 & 2.2e+02 & 7.73e+00\\
10 & 100 & 2 & 222 & 160 & 3.6e+04 & 159 & 1.9e+03 & 1.89e+01\\
10 & 100 & 5 & 381 & 238 & 9.0e+04 & 171 & 2.1e+03  & 4.29e+01\\
10 & 100 & 9 & 589 & 341 & 2.0e+05 & 179 & 2.2e+03 & 9.09e+01\\
10 & 1000 & 2 & 2292 & 1645 & 3.8e+06 & 1587 & 1.9e+04 & 2.00e+02\\
10 & 1000 & 5 & 3832 & 2413 & 9.3e+06 & 1710 & 2.1e+04 & 4.43e+02\\
10 & 1000 & 9 & 5887 & 3440 & 2.0e+07 & 1786 & 2.1e+04 & 9.52e+02\\
10 & 10000 & 2 & 22780 & 16389 & 3.7e+08 & 15832 & 1.9e+05 & 1.95e+03\\
10 & 10000 & 5 & 38512 & 24254 & 9.3e+08 & 17102 & 2.1e+05 & 4.43e+03\\
10 & 10000 & 9 & 59140 & 34567 & 2.0e+09 & 17855 & 2.1e+05 & 9.52e+03\\
100 & 10 & 20 & 88 & 20 & 1.8e+03 & 20 & 2.0e+03 & \textbf{9.00e-01}\\
100 & 10 & 50 & 234 & 77 & 1.8e+04 & 20 & 2.1e+03 & 8.57e+00\\
100 & 10 & 99 & 473 & 194 & 9.2e+04 & 20 & 2.1e+03 & 4.38e+01\\
100 & 100 & 20 & 1104 & 561 & 6.2e+05 & 188 & 2.0e+04 & 3.10e+01\\
100 & 100 & 50 & 2559 & 1284 & 3.3e+06 & 194 & 2.0e+04 & 1.65e+02\\
100 & 100 & 99 & 5067 & 2535 & 1.3e+07 & 196 & 2.0e+04 & 6.50e+02\\
100 & 1000 & 20 & 11391 & 6154 & 7.0e+07 & 1868 & 1.9e+05 & 3.68e+02\\
100 & 1000 & 50 & 26425 & 13666 & 3.6e+08 & 1930 & 1.9e+05 & 1.89e+03\\
100 & 1000 & 99 & 50928 & 25916 & 1.3e+09 & 1959 & 2.0e+05 & 6.50e+03\\
100 & 10000 & 20 & 114003 & 61960 & 7.1e+09 & 18676 & 1.9e+06 & 3.74e+03\\
100 & 10000 & 50 & 264204 & 137056 & 3.6e+10 & 19294 & 2.0e+06 & 1.80e+04\\
100 & 10000 & 99 & 509695 & 259799 & 1.3e+11 & 19577 & 2.0e+06 & 6.50e+04\\
1000 & 10 & 200 & 806 & 88 & 7.1e+04 & 21 & 2.1e+04 & 3.38e+00\\
1000 & 10 & 500 & 2183 & 629 & 1.4e+06 & 21 & 2.1e+04 & 6.67e+01\\
1000 & 10 & 999 & 4541 & 1778 & 8.1e+06 & 21 & 2.1e+04 & 3.86e+02\\
1000 & 100 & 200 & 9781 & 4450 & 4.4e+07 & 198 & 2.0e+05 & 2.20e+02\\
1000 & 100 & 500 & 24787 & 11947 & 3.0e+08 & 200 & 2.0e+05 & 1.50e+03\\
1000 & 100 & 999 & 50039 & 24572 & 1.2e+09 & 200 & 2.0e+05 & 6.00e+03\\
1000 & 1000 & 200 & 100963 & 50491 & 5.1e+09 & 1976 & 2.0e+06 & 2.55e+03\\
1000 & 1000 & 500 & 249909 & 124959 & 3.1e+10 & 1990 & 2.0e+06 & 1.55e+04\\
1000 & 1000 & 999 & 499209 & 249607 & 1.3e+11 & 1994 & 2.0e+06 & 6.50e+04\\
1000 & 10000 & 200 & 1015910 & 512465 & 5.2e+11 & 19756 & 2.0e+07 & 2.60e+04\\
1000 & 10000 & 500 & 2514665 & 1261836 & 3.2e+12 & 19886 & 2.0e+07 & 1.60e+05\\
1000 & 10000 & 999 & 5001083 & 2505044 & 1.3e+13 & 19936 & 2.0e+07 & 6.50e+05\\
	\hline
\end{tabular}
\end{table}